\documentstyle[12pt]{article}
\begin{document}
\thispagestyle{empty}
\vspace*{0.2in}
\begin{flushright}
\hfill\vbox{\hbox{UA/NPPS-1-04}}
\end{flushright}
\vspace{2cm}
\begin{center}
{\large{\bf Non-perturbative corrections to perturbative results\\
pertaining to a four-point Sudakov process in QCD}}\\

\vspace{1.5cm}
A. I. Karanikas$^\dagger$, C. N. Ktorides$^\ddagger$\\
\smallskip
\textit{University of Athens, Physics Department\\
Nuclear \& Particle Physics Section\\
Panepistimiopolis, Ilisia GR 157--71, Athens, Greece}\\
\end{center}

\vspace{3.5cm}
\begin{abstract}
Within the framework of the Fock-Feynman-Schwinger (equivalently, worldline) casting of 
QCD a four-point process, mediated by a closed quark loop, is studied in the Sudakov 
kinematical region, first perturbatively and subsequently with the inclusion of
non-perturbative contributions through the use of the stochastic	vacuum model. Deformations
of the, first order, perturbative results are determined. Finally, the resummation
of the leading logarithmic terms is performed and physical implications of the result are discussed.
\end{abstract}

\vspace*{2cm}

--------------------------------------------------

$^\dagger$e-mail address: akaranik@cc.uoa.gr

$^\ddagger$e-mail address: cktorid@cc.uoa.gr

\newpage

{\bf 1. Introductory remarks}

\vspace*{.2cm}

Sudakov behavior was originally formulated in connection with a vertex function in QED [1].
It pertains to a situation where
a very large momentum transfer is imparted on the fermionic line by a photon and accounts 
for the suppression of probability that photon radiative emission is
not detected above a given (infrared) scale. 
One of its significant implications for pQCD pertains to
exclusive processes and, in particular, to the extent by which perturbative factorization
holds for that corner of phase space where the momentum of the hadron
is basically carried by a single constituent quark (so called Feynman picture). This corresponds, in effect, to a 
low-x situation, given that it implies the vanishing of the momentum fraction carried by the
other, accompanying, valence quark(s). In the perturbative context the aformentioned infrared scale should,
of course, be possibly close, but certainly above $\Lambda_{QCD}$. In practice, it corresponds to the 
(inverse) separation scale between the participating and the low-x quark(s).
The central objective of this work is to study  possible modifications
induced by non-perturbative contributions to the result of Refs. 2-4.

The methodological tools for our analysis are provided by: a) The 
Fock[5]-Feynman [6]-Schwingwer [7] (FFS), also known as worldline, casting 
of QCD [8-10], on account of certain advantages it offers in connection 
with the application of eikonal approximation methods in QCD [11] and b) the 
stochastic vacuum model (SVM) [12-16], which has been constructed for the 
purpose of confronting nonperturbative issues in QCD. This scheme 
is founded on three basic  axioms whose aim is to define the vacuum structure 
of the theory, several consequences of which have been succesfully
checked against lattice results [17]. 

In this work we address the Sudakov suppression issue in an `ìdealized' context which pertains to the
four-point process depicted in Fig. 1. Once isolating the kinematical corner
which ensures Sudakov kinematics for the process, namely large momentum transfer at {\it each
one of the four vertices}, we proceed to study the problem
concerning the extent to which
non-perturbative contributions affect the relevant, Sudakov result.

Following a brief exposition of the FFS-worldline formalism regarding relevant basic tools,
we shall proceed, in Section 3, to determine the Sudakov 
kinematical region, as it pertains to the four-point process under
discussion and study its perturbative implications. Section 4
is devoted to the quantitative study 
of lowest order, non-perturbative contributions to the four-point  
process (always in the Sudakov kinematical region) and determine the modifications induced by them.
Once this is done a resummation of the overall soft contributions will be carried out (section 5),
which will facilitate direct comparisons with the pQCD result [2]. 
Concluding remarks will be presented in the last section. Finally, the 
technical manipulations involved in the proof of a set of equations
in the main text are relegated to an  Appendix.

\vspace*{.3cm}

{\bf 2. General formalism}

\vspace*{.2cm}

Consider the n-point Green's function entering a QCD process defined by a
product of local currents
$J_\mu(x)\equiv \bar{\psi}(x)\Gamma_\mu\psi(x)$, where $\Gamma_{\mu}$ is an appropriate 
element of the Dirac-Clifford algebra. One writes 
\begin{eqnarray}
&&G_{\mu_n\cdot\cdot\cdot\mu_1}(x_n,\cdot\cdot\cdot,x_1)=\langle J_{\mu_1}(x_n)
\cdot\cdot\cdot J_{\mu_n}(x_1)\rangle_{\psi,A}\nonumber\\&&\quad
=Tr_C\langle iG(x_1,x_n|A)\Gamma_{\mu_n}iG(x_n,x_{n-1}|A)\Gamma_{\mu_{n-1}}
\cdot\cdot\cdot iG(x_2,x_1|A)\Gamma_{\mu_1}\rangle_A.
\end{eqnarray}
In the FFS-worldline formulation the propagators entering the above relation are given by [8-10]
\begin{eqnarray}
iG(x,x'|A)&\equiv&\int_0^\infty dT\,e^{-m^2T}\int\limits_{x(0)=x',\,x(T)=x}{\cal D}x(t)
e^{-{1\over 4}\int_0^Tdt\dot{x}^2(t)}[m-{1\over 2}\gamma\cdot\dot{x} (T)]
\nonumber\\&&\quad\times\Phi[\dot{x}]^{[1/2]} 
{\rm P} exp\left(ig\int_0^Tdt\dot{x}\cdot A\right).
\end{eqnarray}
In the above relation $\Phi^{[1/2]}$ enters as a factor which accounts for the spin 
of the propagating particle modes. More specifically, spin effects are represented by the term
${\rm exp}\left({i\over 2}\int_0^Tdt\,\sigma_{\mu\nu}F_{\mu\nu}\right)$, which should
be positioned inside the path ordered Wilson exponential. It is possible, on the other hand,
to factorize the spin effects via a partial integration [17] and embody them on the so-called 
{\it spin factor} [18] which is given by 
\begin{equation}
\Phi[\dot{x}]^{[1/2]}\equiv{\rm P}exp\left({i\over 2}\int_0^Tdt\,\sigma_{\mu\nu}\omega_{\mu\nu}[\dot{x}]\right),
\end{equation}
where $\omega_{\mu\nu}={T\over 2}[\ddot{x}_\mu(t)\dot{x}_\nu(t)
-\ddot{x}_\nu(t)\dot{x}_\mu(t)]$ is a `torsion' tensor and $\sigma_{\mu\nu}$ are
the spin-1/2 generators for the Lorentz group\footnote{By abuse of language the
characterization `Lorentz group' accomodates the Euclidean case as well.}.
In the context of the eikonal approximation
where spin effects are absent, it can be set to unity. On the other hand
its presence will become important in connection with the non-perturbative considerations 
that will eventually enter our analysis, at which point it will be explicitly taken into account.

Under suitable parametrizations the four-point Green's function,
to which we specialize our considerations from hereon, can be cast
in the following form (note that the $s_i,\,i=1,\cdot\cdot\cdot, 4$ serve
to parametrize the, closed, quark contour -see Fig. 1- with $T=s_4$ being the
`total time')
\begin{eqnarray}
&&G_{\mu_4\cdot\cdot\cdot\mu_1}=\left[\prod\limits_{i=4}^1\int_0^\infty 
ds_i\,\theta(s_i-s_{i-1})e^{-m^2(s_i-s_{i-1})}\right]
\int\limits_{z(0)=z(T)=x_1}{\cal D}z(t)e^{-{1\over 4}\int_0^Tdt\dot{z}^2(t)}
\nonumber\\&&\quad\times\left[\prod\limits_{i=4}^2\delta(z(s_{i-1})-x_i)\right]
S_{\mu_4\cdot\cdot\cdot\mu_1}[\dot{z}]\left
\langle Tr_C{\rm P} {\rm exp}\left[ig\int_0^Tdt\dot{z}\cdot A\right]\right\rangle_A,
\end{eqnarray}
where
\begin{equation}
S_{\mu_4\cdot\cdot\cdot\mu_1}[\dot{z}]\equiv\prod\limits_{i=4}^2[m-{1\over 2}
\gamma\cdot\dot{z}_i(s_i)]\Gamma_{\mu_i}\Phi^{[1/2]}.
\end{equation}
 
Going to momentun representation one obtains, after taking into consideration
momentum conservation around the fermionic loop, the following expression 
for the Green's function
\begin{eqnarray}
&&\bar{G}_{\mu_4\cdot\cdot\cdot\mu_1}(p_4,\cdot\cdot\cdot,p_1)
=\left[\prod\limits_{i=4}^1\int_0^\infty 
ds_i\,\theta(s_i-s_{i-1})e^{-m^2(s_i-s_{i-1})}\right]
\int\limits_{z(0)=z(T)=0}{\cal D}z(t){\rm exp}\left[-{1\over 4}\int_0^Tdt\dot{z}^2(t)\right.
\nonumber\\&&\quad \left. -i\sum\limits_{i=1}^{3}p_i\cdot z(s_i)\right]
S_{\mu_4\cdot\cdot\cdot\mu_1}[\dot{z}]
\left\langle Tr_C{\rm P} {\rm exp}\left[ig\int_0^Tdt\dot{z}\cdot A\right]\right\rangle_A.
\end{eqnarray}

Turning our attention to the expectation value of the Wilson exponential let us
introduce the quantity $C[z]$ by
\begin{equation}
{\rm exp}C[z]\equiv\left\langle {\rm Pexp}\left[ig\int_0^Tdt\dot{z}\cdot A\right]\right\rangle_A.
\end{equation}
To the extent that $C[z]$ is expected to condense the (soft) dynamics of the system,
one should develop a strategy for its computation. To this end, we adopt the cumulant expansion,
originally proposed in Ref [20] in connection with stochastic processes and whose significance
for field theoretical applications has been extensively employed, see, e.g., Refs [12-16, 21].
Denoting cummulant correlators with double brackets, one writes
\begin{equation}
{\rm exp}\,C[z]=\exp
\left[\sum\limits_{r=1}^\infty \frac{(ig)^r}{r!}
\int_0^T dt_r\cdot\cdot\cdot
\int_0^T dt_1\dot{z}_{\mu_r}(t_r)\cdot\cdot\cdot\dot{z}_{\mu_1}(t_1)\langle
\langle A_{\mu_r}(t_r)\cdot\cdot\cdot A_{\mu_1}(t_1)\rangle\rangle\right],
\end{equation}
which reduces to an identity for the case when each term in the sum finite. 

The cumulant expansion obeys a set of
recursive relations in terms of field theoretical correlators, having
the form (Lorentz and color indices for the vector field omitted)
\begin{eqnarray}
&&\langle\langle A(t)\rangle\rangle=\langle A(t)\rangle,\nonumber\\&&
\langle\langle A(t_1)A(t_2)\rangle\rangle=\langle P(A(t_1)A(t_2))\rangle
-\langle A(t_2)\rangle\langle A(t_1)\rangle,\,{\rm etc.}
\end{eqnarray}

For SVM applications, it becomes important to note that factorization rules axiomatically
operate for higher $(\geq 3)$ point gluon field strength correlators. As a final note, 
let us remark that, for the perturbative case, Eq (6) corresponds
to a formal relation which needs to be regularized by one method or other before 
it aquires a concrete meaning. Once this is done a regularization mass appears, which
induces the application of appropriate
renormalization group equations.

The expression for the four-point Green's function by can now be summarized as follows
\begin{equation}
\bar{G}_{\mu_4\cdot\cdot\cdot\mu_1}=\left[\prod\limits_{i=4}^1\int_0^\infty 
ds_i\,\theta(s_i-s_{i-1})e^{-m^2(s_i-s_{i-1})}\right]\int\limits_{z(0)=z(T)=0}{\cal D}z(t)
S_{\mu_4\cdot\cdot\cdot\mu_1}[\dot{z}]Tr_C e^{-I[z]},
\end{equation}
where
\begin{equation}
I[z]={1\over 4}\int_0^Tdt\dot{z}^2(t)+i\sum\limits_{i=1}^{3}p_i\cdot z(s_i)-C[z].
\end{equation}

The quantitative analysis in this paper will be conducted
with reference to the above generic expression, which can be seen as playing the
role of an `action functional' for a `particle living on the loop'. The 
complexity of the problem is, clearly, determined by the `interaction' term $C[z]$. 
Given our intention to extend our considerations to
both the perturbative and the non-perturbative domain
of QCD the quantity $C[z]$, which isolates all the dynamics in the `particle action
functional', will be organized as follows  
\begin{equation}
C[z]=C_{{\rm pert}}[z] +C_{{\rm non-pert}}[z]+ C_{{\rm interf}}[z].
\end{equation}

Our objective is to apply the particle-based casting of QCD described above by 
focusing on long distance effects. To this end we shall employ semiclassical,
eikonally-based techniques. Consider, in this connection,
the variation of the `action functional' $I[z]$ entering Eq. (7) with respect 
to $z_\mu(t)$. One writes
\begin{equation}
\frac{\delta I[z]}{\delta z_\mu (t)}=-{1\over 2}\ddot{z}_\mu(t)+i\sum\limits_{i=1}^{3}
p_{i\mu}\delta(t-s_i)-\frac{\delta}{\delta z_\mu(t)}C[z].
\end{equation}
Its stationary points furnish `classical' equations of motion the integral form
of which reads
\begin{equation}
z_\mu^{\rm cl}(t)= z_\mu^0(t)+\int_0^Tdt'K_\mu[t,t';z_\mu^{{\rm cl}}(t')].
\end{equation}
with the integral kernel reading as follows
\begin{equation}
K_\mu[t,t';z_\mu^{{\rm cl}}(t')]=- 2\Delta(t,t') \frac{\delta C[z^{{\rm cl}}]}{\delta z_\mu^{{\rm cl}} (t')}.
\end{equation}

In the above equation $z_\mu^0(t)$ enters as the solution of the linear part of the
`system' and is given by 
\begin{equation}
z_\mu^0(t)=2i\sum\limits_{i=1}^{3}p_{i\mu}\Delta(t,s_i),
\end{equation}
where
\begin{equation}
\Delta(t,t') \equiv \frac{t(T-t')}{T} \theta(t'-t)+\frac{t'(T-t)}{T}\theta(t'-t).
\end{equation}
and corresponds to a straight line trajectory.

Our expectation is that  we can get a good estimate
of the final result via the appplication of an iterative procedure 
on Eq (14), given the presence of a large scale, furnished by the incoming energy, on which an asymptotic 
calculation can be based. From a geometrical standpoint the classical
solution given by Eq (14) has a profile which can be characterized 
as being of an `eikonal type'. Our whole effort in this paper basically amounts
to assessing first order contributions from the iterations coming both
from the perturbative and the background (non-perturbative) gauge field sector.
 
For a straight (world)line segment, joining two consecutive vertices 
specified, e.g, by the interval $[s_{k-1},s_k)$ in fig. 1 one writes 
($\bar{s}_k$ denotes its midpoint) 
\begin{equation}
\sum\limits_{i=1}^{3}p_{i\mu}\dot{\Delta}(t_k,s_i)=
\sum\limits_{i=1}^{3}p_{i\mu}\dot{\Delta}(\bar{s}_k,s_i)\equiv q_{k\mu}
\end{equation}
with $t_k\in[s_{k-1},s_k)$. 
Accordingly, the corresponding equations of motion read
\begin{equation}
z_\mu^0=2iq_{k\mu}(t_k-s_{k-1})+z_\mu^0(s_{k-1}),\quad z_\mu^0(0)=z_\mu^0(T)=0.
\end{equation}
The above, `zero' order solutions of the `equations of motion', 
describe quark `propagation' on straight-line (eikonal) contours. In the following section
we shall use them as input solutions for initiating first order perturbative contributions to the
four-point Green's function in the Sudakov kinematical region. 

\vspace{.3cm} 

{\bf 3. Perturbative contributions to self-energy and vertex corrections -Sudakov behavior}

\vspace*{.2cm}

In this section we shall, with reference to the four-point Green's function, consider
perturbative correction stemming from: (a) virtual
gluon emission and absorbtion off an (eikonal) fermionic line and (b) virtual gluon
exchanges between such lines accross a given vertex. The no-recoil approximation associated
with our restriction to straight line segments is equivalent to the `absence' of gluon radiation.
The kinematical region corresponding to large momentum transfer across a given vertex
produces the conditions which give rise to the Sudakov form factor [1]. Going a step further, we
shall determine the kinematical region for a high energy, four-point process which secures Sudakov conditions 
for each vertex. This will lead to Sudakov suppression behavior that is
consistent with well known results [2-4].

\vspace{.2cm} 

{\bf 3a. Perturbative self-energy and vertex corrections}

\vspace*{.1cm}

In the perturbative sector the expression for $C[z]$,  
to second order of the coupling constant, is given by  
\begin{equation}
C_{{\rm pert}}^{(2)}[z]=-{1\over2}g^2C_F\frac{\mu^{4-D}}{4\pi^{D/2}}\Gamma\left({D\over 2}-1\right)
\int_0^Tdt_1\int_0^Tdt_2\frac{\dot{z}(t_2)\cdot\dot{z}(t_1)}{|z(t_2)- z(t_1)|^{D-2}}.
\end{equation}
Accordingly, the following second order expression is determined for the integral kernel: 
\begin{eqnarray}
K^{(2)}_{\mu,{\rm pert}} &=& g^2C_F\frac{\mu^{4-D}}{4\pi^{D/2}}2(D-2)\Delta(t,t')
\int_0^Tdt''{1\over |z^{{\rm cl}}(t'')-z^{{\rm cl}}(t')|^D}\nonumber\\&&\times
[\dot{z}^{{\rm cl}}(t'')\cdot \dot{z}^{{\rm cl}}(t')(z_\mu^{{\rm cl}}(t'')-z_\mu^{{\rm cl}}(t'))
-\dot{z}^{{\rm cl}}(t'')\cdot \dot{z}^{{\rm cl}}(t')(z_\mu^{{\rm cl}}(t'')-z_\mu^{{\rm cl}}(t'))].
\end{eqnarray}

Generally speaking, $K^{(2)}_{\mu,{\rm pert}}$ will induce, through an iterative
procedure applied to Eq (14), the deviation of the classical solution
from the straight line configuration. For the purposes of the present section,
in which we shall remain within the perturbative domain of QCD, 
only leading order corrections will be considered. This, as has already been mentioned,
amounts to employing the eikonal approximation. 

Consider the quantity
\begin{equation}
I_{k-1,k}\equiv{1\over2}g^2C_F\frac{(4\mu^2)^{2-D/2}}{4\pi^{D/2}}\Gamma\left({D\over 2}-1\right)
\int_{s_{k-1}}^{s_k}dt_2\int_{s_{k-1}}^{s_k}dt_1
\frac{\sum\limits_{i,j=1}^{3}p_i\cdot p_j\dot{\Delta}(t_2,s_i)\dot{\Delta}(t_1,s_j)}
{|\sum\limits_{i=1}^{3}p_i\left(\Delta(t_2,s_i)-\Delta(t_1,s_i)\right)|^{D-2}},
\end{equation}
which describes, to first perturbative order, a self energy insertion to the quark worldline
path segment between vertices $k-1$ and $k$. 
Referring to Eq (18) one finds, after some simple algebra, that the denominator 
inside the integral equals $q_{k\mu}(t_2-t_1)$.
This leads to the result
\begin{eqnarray}
I_{k-1,k} &=&{1\over2}g^2C_F\frac{(4\mu^2)^{2-D/2}}{4\pi^{D/2}}\Gamma\left({D\over 2}-1\right)
\int^{s_k-s_{k-1}}_0 dt_2\int^{s_k-s_{k-1}}_0 dt_1\frac{1}{|t_2-t_1|^{D-2}}\nonumber\\
&=&-{\alpha_S\over 2\pi}C_F{1\over\varepsilon}-{\alpha_S\over 2\pi}C_F
{\rm ln}\left(4\pi e^{-\gamma_E+2}\mu^2L_k^2\right)+{\cal O}(\varepsilon),
\end{eqnarray}
where $\gamma_E$ is the Euler-Mascheroni constant, $\varepsilon=2-D/2$ and
$L_k^2\equiv q_k^2(s_k-s_{k-1})^2$.

Turning our attention to gluon exchanges between adjacent quark lines,
accross a given vertex, we find ourselves having to deal with the quantity
\begin{equation}
I_{k-1\,\,\,\,k+1}^{\quad k}\equiv{1\over2}g^2C_F\frac{(4\mu^2)^{2-D/2}}{4\pi^{D/2}}\Gamma
\left({D\over 2}-1\right)\int_{s_{k-1}}^{s_k}dt_1\int_{s_{k}}^{s_{k+1}}dt_2
\frac{\sum\limits_{i,j=1}^{3}p_i\cdot p_j\dot{\Delta}(t_2,s_i)\dot{\Delta}(t_1,s_j)}
{|\sum\limits_{i=1}^3 p_i\left(\Delta(t_2,s_i)-\Delta(t_1,s_i)\right)|^{D-2}}.
\end{equation}
For subsequent purposes we also introduce the quantities
\begin{equation}
c_{k,k+1}\equiv\frac{q_k\cdot q_{k+1}}{|q_k||q_{k+1}|}.
\end{equation}

As shown in the Appendix, the $I_{k-1\,\,\,\,k+1}^{\quad k}$ can be brought into the following form:
\begin{equation}
I_{k-1\,\,\,\,k+1}^{\quad k}={1\over\varepsilon}{\alpha_S\over2\pi}C_F
V_4(c_{k,k+1})+{\alpha_S\over2\pi}C_F V_4(c_{k,k+1})
{\rm ln}\left(4\pi e^{-\gamma_E+2}\mu^2L_k^2\right)+{\alpha_S\over2\pi}C_F
f(c_{k,k+1})+{\cal O}(\varepsilon),
\end{equation}
where\footnote{For notational economy we let $c$ stand for the generic $c_{k,k+1}$.}
\begin{equation}
V_4(c)={1\over 2}\frac{c}{\sqrt{1-c^2}}{\rm arctg}\frac{\sqrt{1-c^2}}{c}
={1\over 2}\theta{\rm cot}\theta,\quad {\rm cos}\theta\equiv c
\end{equation}
with $\theta$ the angle, in Euclidean space, formed by the quark worldline
paths meeting at the vertex $k$ and where
\begin{equation}
f(c)\equiv \lim_{\varepsilon\rightarrow 0}{1\over\varepsilon}[V_{4-2\varepsilon}(c)
-V_4(c)].
\end{equation}

Going over to Minkowski space-time entails the transcription $\theta\rightarrow-i\gamma$,
which gives $\theta{\rm cot}\theta\rightarrow\gamma{\rm coth}\gamma$. Denoting the 
total perturbative contribution from self energy (se) and vertex (v) corrections by 
$I^{{\rm se,v}}_k(\equiv I_{k-1,k} + I_{k-1\,\,\,\,k+1}^{\quad k} + I_{k+1\,\,\,\,k-1}^{\quad k})$, 
one obtains
\begin{equation}
I^{{\rm se,v}}_k=
{\alpha_S\over 2\pi}C_F(\gamma_{k,k+1}{\rm coth}\gamma_{k,k+1}-1)\left({1\over\varepsilon}
+{\rm ln}(4\pi e^{-\gamma_E})\right)
+{\alpha_S\over2\pi}c_F(\gamma_{k,k+1}{\rm coth}\gamma_{k,k+1}-1) 
{\rm ln}\frac{\mu^2}{\lambda_k^2},
\end{equation}
where
\begin{equation}
{1\over\lambda_k^2}\equiv L_k^2 exp\left[2+2
\frac{f(\gamma_{k,k+1})}{\gamma_{k,k+1}{\rm coth}\gamma_{k,k+1}-1}\right].
\end{equation}

Consider, now, the case for which
\begin{equation}
\frac{|q_k\cdot q_{k+1}|}{|q_k||q_{k+1}|} =\cosh\gamma_{k,k+1}\gg1,
\end{equation}
i.e. where a large momentum is imparted at the vertex. Such a situation gives rise,
under the circumstance that gluon radiation is unobserved, to the Sudakov form factor.  
As already mentioned, the aformentioned unobservability is accounted for by the smooth (eikonal type)
spinorial lines entering and leaving the vertex. This is consistent with the obervation that, in the
perturbative context, the emission of a gauge field quantum from 
off the worldline contour of a matter particle induces a point of derivative discontinuity [18],
whereas we have been restricting our considerations to smooth worldline contours. 
In passing we note that the kinematical limit specified by Eq (31) leads to
\begin{equation}
f(\gamma_{k,k+1})\simeq\frac{1}{4} \gamma_{k,k+1}^2+\frac{1}{2}\gamma_{k,k+1} 
\ln\left(\frac{L_{k+1}}{L_k}\right)- \gamma_{k,k+1},
\end{equation}
which in turn gives
\begin{equation}
\frac{1}{\lambda^2_k}\simeq L_{k+1}{L_k}\exp\left(\frac{1}{2} \gamma_{k,k+1}\right)
\simeq L_{k+1}{L_k}\left(2\frac{|q_k\cdot q_{k+1}|}{|q_k||q_{k+1}|}\right)^{1/2}.
\end{equation} 

From the above follows that that the finite part of the vertex function, cf. Eq (29),
assumes the form
\begin{equation}I_k^{{\rm se,v}}\simeq\frac{\alpha_S}{2\pi}C_F\ln\left(2
\frac{|q_k\cdot q_{k+1}|}{|q_k||q_{k+1}|}\right) \ln\left[\mu^2\left(2
 \frac{|q_k\cdot q_{k+1}|}{|q_k||q_{k+1}|}\right)^{1/2}L_{k+1}{L_k}\right]. 
\end{equation}

Of course, what we are presently 
interested in is not the kinematical region pertaining to a single vertex function but the
one associated with the four-point process as a whole. This means that
the sought for kinematical region for the four-point process is one where the Sudakov
conditions prevail for {\it each single vertex}. This is the task that we shall undertake in the
following subsection.

\vspace{.2cm} 

{\bf 3b. Sudakov suppression for a four-point process in pQCD}

\vspace*{.1cm}

Consider the four-point process depicted in fig 1. We introduce parameters $x_i,\,i=1,2,3,4$ as follows
\begin{equation}
x_1\equiv\frac{s_1}{T},\,x_2\equiv\frac{s_2-s_1}{T},\,x_3\equiv\frac{s_3-s_2}{T} ,\,x_4\equiv\frac{s_4-s_3}{T}
\end{equation}
on the basis of which one determines (see fig 1 for designations of the momenta)
\begin{equation}
q_1=(1-x_1)p_A-x_2(p_{B'}-p_B)+x_3p_B,
\end{equation}
while the rest of the $q_i,\,i=2,3,4$ are specified by the relations
\begin{equation}
q_2=q_1-p_{A'},\quad q_3=q_1-(p_{A}+p_B), \quad q_4=q_1-p_{A}.
\end{equation}

What is {\it a priori} given is that the incoming energy $s=(p_A+p_B)^2$ sets the large scale 
for our problem. Let us then take $p_A^2=p_B^2=p^2_{A'}=p^2_{B'}=M^2\ll s^2$, which means that 
$p_A\cdot p_B=p_{A'}\cdot p_{B'}\simeq s/2$,
$p_A\cdot p_{A'}=p_B\cdot p_{B'}\simeq -t/2={1\over 2}s(1-cos\theta)$ ($\theta$ the scattering angle 
in the cm frame) and $p_{A'}\cdot p_B=p_A\cdot p_{B'}\simeq -{u/2}$. 
With these specifications one can easily 
deduce that the conditions of Eq (31) for $k=1,2,3,4$ cannot be met 
unless three of the $x_i$ are very small. 
Introducing a mass scale $m^2\ll s$ the
desirable kinematical region can be achieved by setting
$x_2\simeq x_3\simeq x_4\simeq{m^2/s}$, $x_1\simeq 1$.
Simple algebraic manipulations show that for such a choice 
\begin{equation}
q_1^2\sim \frac{m^4}{s},\,q_2^2\sim m^2,\,q_3^2\sim s,\,q_4^2\sim m^2
\end{equation}
and
\begin{equation}
|q_1\cdot q_2|\sim m^2,\,|q_2\cdot q_3|\sim s,\,|q_3\cdot q_4|\sim s,\,|q_4\cdot q_1|\sim m^2
\end{equation}
from which one trivially verifies that a large momentum transfer takes place at each on of the 
four vertices. Explicitly, one deduces that
\begin{equation}
\cosh\gamma_{12} \sim \cosh\gamma_{23} \sim \cosh\gamma_{34} \sim \cosh\gamma_{41} 
\sim\left({s\over m^2}\right)^{1/2}.
\end{equation}

With the kinematics in place, one is in position to commence with the computation 
of the amplitude. Consider, first, the contribution from the free part of the
`particle action functional' pertaining to the four-point contour. One obtains
\begin{eqnarray}
&&I_{free}^{(0)}={1\over 4}\int_0^Tdt(\dot{z}^0(t))^2+i\sum\limits_{i=1}^3p_i\cdot z^0(s_i)
=\sum\limits_{i=1}^4\sum\limits_{j=i+1}^4 p_ip_jG(s_i,s_j)\nonumber\\&&
\simeq{s\over2} [G(s_1,s_2)+G(s_3,s_4)]+{t\over2} [G(s_2,s_3)+G(s_1,s_4)]
+{u\over2} [G(s_2,s_3)+G(s_2,s_4)],
\end{eqnarray}
where $G(s_i,s_j)={1\over T}|s_i-s_j|(T-|s_i-s_j|)$ and where we have adopted the convention
$p_1=p_{A'},\,p_2=p_{B'},\,p_3=-p_{B},\,p_4=-p_A$.
For the kinematical region under consideration one finds
\begin{equation}
I_{free}^{(0)}\sim Tm^2.
\end{equation}
This behavior can be interpreted as signifying the presence of an infrared cutoff for contours with
length larger than $1/m$ and can thereby be used as the defining relation for the scale $m^2$.

Taking into consideration the above specifications Eq (29) assumes the following form
\begin{equation}
I^{{\rm se,v}}_k\simeq {\alpha_s\over 2\pi}C_F {\rm ln}\left(\frac{s}{m^2}\right)^{{1\over 2}}\left\{
\ln\left[\frac{\mu^2}{m^2}\frac{|q_k||q_{k+1}|}{m^2}\left(\frac{s}{m^2}\right)^{{1\over 4}}
\right]+ 2\ln(Tm^2)+\ln(x_kx_{k+1})\right\}.
\end{equation}
Summing contributions from all four vertices one finally obtains
\begin{equation}
I^{{\rm se,v}}\simeq {\alpha_s\over 2\pi}C_F {\rm ln}\left(\frac{s}{m^2}\right)\left\{2
\ln\left[\frac{\mu^2}{m^2}\left(\frac{s}{m^2}\right)^{{1\over 4}}
\right]+ 4\ln(Tm^2)+\ln(x_1x_2x_3x_4)\right\}.
\end{equation}

Going, now, to the amplitude we note the following. First, integrations 
over the parameters $x_2$, $x_3$, $x_4$ in the 
region $x_2\sim x_3\sim x_4\sim m^2/s$ should be performed and second, the computed (up
to first perturbative order) contributions to the `particle action functional' enter as exponentials. We
proceed with their assessment by considering one by one the three terms composing 
$I^{{\rm se,v}}$.
 
The contribution coming from the last term of the above equation gives
rise to integrals of the form
\begin{equation}
\int_0^{m^2/s}dx_kx_k^{-{\alpha_S\over2\pi}C_F\ln\left({s\over m^2}\right)}
=\frac{1}{1-{\alpha_S\over2\pi}C_F\ln\left({s\over m^2}\right)}
\left({s\over m^2}\right)^{-\left\{1-{\alpha_S\over2\pi}C_F\ln\left({s\over m^2}\right)\right\}}
\end{equation}
ensuring that the kinematical region under consideration {\it does not} have zero measure.
More specifically, as will be argued in the next section, the strong coupling constant can be
defined at a scale for which
\begin{equation}
{\alpha_S\over2\pi}C_F\ln\left({s\over m^2}\right)\sim{\cal O}(1),
\end{equation}
hence it should eventually give non-negligible contributions.

The second term gives 
\begin{equation}
{\rm exp}\left[-{\alpha_s\over2\pi}C_F{\rm ln}\left(\frac{s}{m^2}\right)4{\rm ln}(Tm^2)\right]=
(Tm^2)^{-4{\alpha_s\over2\pi}C_F{\rm ln}\frac{s}{m^2}}.
\end{equation}
Integration over $T$, as entailed by the worldline expression, leads to a final contribution
of the form
\begin{eqnarray}
\int_0^\infty dT\,T^3 (Tm^2)^{-4{\alpha_S\over2\pi}C_F{\rm ln}\frac{s}{m^2}}e^{-Tm^2}&\sim&
\int_0^\infty dT\, (T)^{-1+4\left(1-{\alpha_S\over2\pi}C_F{\rm ln}\frac{s}{m^2}\right)}e^{-Tm^2}
\nonumber\\&&\sim\Gamma\left[4\left(1-{\alpha_S\over2\pi}C_F{\rm ln}
\frac{s}{m^2}\right)\right],
\end{eqnarray}
which is of finite order.

The notable contribution at the perturbative level is furnished by the first term, i.e.
\begin{equation}
\bar{G}\sim {\rm exp}\left\{-{\alpha_S\over \pi}C_F {\rm ln}\left(\frac{s}{m^2}\right)
{\rm ln}\left[\frac{\mu^2}{m^2}\left(\frac{s}{m^2}\right)^{{1\over 4}}
\right]\right\}
\end{equation}
and displays Sudakov suppression for an exclusive, high energy
four point process.

As a final note, let us observe that the presence of the regularization mass scale $\mu$ 
implies an eventual renormalization group
running, given the final result should not depend on it. Such a running calls for `initial conditions'
set at some minimum scale $\mu_{min}$. We shall defer the discussion of this matter until 
non-perturbative deformations to the integral kernel will have also been taken into account.

\vspace*{.3cm}

{\bf 4. Non-perturbative considerations: Deformation of the straight line contours}

\vspace*{.2cm}

As mentioned in the introduction, non-perturbative effects associated with the four-point
Green's function under study will be assessed through the employment of the Stochastic Vacuum 
Model (SVM) [12,13]. Its central objective is to incorporate established
observations/results regarding the structure of the QCD vacuum whose starting point can be traced
to  Ref [21]. The SVM attempts to summarize all that is known and/or surmized about
the properties of the QCD vacuum through a set of three axioms which are expressed in terms of 
field strength, as opposed to field potential, correlators. The underlying
stochasticity assumption for the vacuum state facilitates the direct application
of the cumulant expansion -see, e.g., relevant review articles [22,16,20]. 
A concrete, as well as practical, way to apply the SVM scheme to specific situations 
is by using the background gauge fixing method [23], with the background
gauge fields becoming the agents of the non-perturbative dynamics. Specifically, 
one employs the gauge potential splitting
$A_\mu^a=\alpha_\mu^a\,+\,B_\mu^a$
where the $\alpha_\mu^a$ are
associated with the usual perturbative field modes. The 
$B_\mu^a$, on the other hand, enter as {\it dynamical} fields, assigned with the task 
of carrying the non-perturbative physics through field strength correlators which
obey factorization rules according to which
higher order gluon field strength correlators
are expressible in terms of two-point ones (third axiom of the SVM). 

Some preliminary matters should be dealt with from the outset. First, let us mention
that we shall be employing throughout the Fock-Schwinger (F-S) gauge [24,7], namely 
\begin{equation}
B_\mu^a(x)=-\int_{x_0}^xdu_\nu(\partial_\mu u_\rho)F_{\rho\nu}^a(u)=
-(x-x_0)_\nu\int_0^1d\alpha\,\alpha F_{\mu\nu}(x_0+\alpha(x-x_0)),
\end{equation}
which facilitates the passage from gauge field potential to field strength correlators.
Specifically, one has
\begin{eqnarray} 
&&\langle gB_{\mu_2}(x(t_2))gB_{\mu_1}(x(t_1))\rangle_B=(x_2-x_0)_{\nu_2}
(x_1-x_0)_{\nu_1}\int_0^1d\alpha_2\alpha_2\int_0^1d\alpha_1\alpha_1\nonumber\\&&\quad\quad\quad
\times\langle gF_{\mu_2\nu_2}^c(x_0+\alpha_2(x_2-x_0))
gF_{\mu_1\nu_1}^c(x_0+\alpha_1(x_1-x_0))\rangle_B.
\end{eqnarray}

Upon setting $u_i=x_0+\alpha_ix(t),\,i=1,2$ one determines
\begin{eqnarray}
&&2tr_C\langle gF_{\mu_2\nu_2}^c(u_2)gF_{\mu_1\nu_1}^c(u_1)\rangle_B\equiv 2N^C
\Delta^{(2)}_{\mu_2\nu_2,\mu_1\nu_1}(u_2-u_1)\nonumber\\&&\quad\quad
=2tr_C\langle\phi(x_0,u_2) gF_{\mu_2\nu_2}^c(u_2)\phi(u_2,x_0) \phi(x_0,u_1) 
gF_{\mu_1\nu_1}^c(u_1)\phi(u_1,x_0) \rangle_B,
\end{eqnarray}
where $\phi(x_0,u_i)={\rm P}exp\left(ig\int_{u_i}^{x_0}dv\cdot B(v)\right)$. In the F-S gauge
this factor is unity. Its insertion serves to underline the gauge invariance of the
field strength correlator.

Consider, now, the second term entering the rhs of Eq (12). Substituting the label `non-pert' by `bkgd',
given that the non-perturbative dynamics of QCD are carried by the gauge
field modes $B_\mu^a$, we write for the main object of computational interest
\begin{eqnarray}
C^{(2)}_{\rm bkgd}&=&-C_F\frac{N_c}{N_c^2-1}\int_0^1d\alpha_2\,\alpha_2\int_0^T
dt_2\dot{z}(t_2)\int_0^1d\alpha_1\,\alpha_1\int_0^Tdt_1\dot{z}(t_1)
\Delta^{(2)}_{\mu_2\nu_2,\mu_1\nu_1}[\alpha_2z(t_2)-\alpha_1z(t_1)]\nonumber\\&&\quad
=-C_F\frac{N_c}{N_c^2-1}\oint d\sigma_{\mu_2\nu_2}(u_2)
\oint d\sigma_{\mu_1\nu_1}(u_1)\Delta^{(2)}_{\mu_2\nu_2,\mu_1\nu_1}(u_2-u_1),
\end{eqnarray}
having shifted the reference point $x_0$ to zero, so that $u_i\equiv\alpha_iz(t_1),\,i=1,2$.

A convenient representation of the correlator, to which we shall be referring 
throughout our analysis, is [22,16,20]
\begin{eqnarray}
&&\Delta^{(2)}_{\mu_2\nu_2,\mu_1\nu_1}(z)= (\delta_{\mu_2\mu_1}
\delta_{\nu_2\nu_1}-\delta_{\mu_2\nu_1}\delta_{\nu_2\mu_1})
D(z^2)\nonumber\\&&+{1\over 2}\frac{\partial}{\partial z_{\mu_1}}\left[(z_{\mu_2}
\delta_{\nu_2\nu_1}-z_{\nu_2}\delta_{\mu_2\nu_1})D_1(z^2)\right]
+{1\over 2}\frac{\partial}{\partial z_{\nu_1}}\left[(z_{\nu_2}
\delta_{\mu_2\mu_1}-z_{\mu_2}\delta_{\nu_2\mu_1})D_1(z^2)\right].
\end{eqnarray}
Now, the integral kernel, in terms of which the lowest order non-perturbative
corrections to the amplitude will be determined, is given by
\begin{equation}
K^{(2)}_{\mu,{\rm bkgd}}[t,t';s^{{\rm cl}}(t')]=-2\Delta(t,t') \frac{\delta}
{\delta z_\mu^{{\rm cl}}(t')}C^{(2)}_{\rm bkgd}[z^{{\rm cl}}]
\end{equation}
so, naturally, we need to focus our attention to $C^{(2)}_{\rm bkgd}[z]$.

Given the large energy scale $s$ entering our problem, we can safely assume that any ratio formed between
some other (energy) scale characterising the system intrinsically, such as the string
tension or the inverse correlation length (squared) and $s$ will be very small.  
It can be demonstrated that, under these circustances 
\begin{equation}
C^{(2)}_{\rm bkgd}[z]\approx -\sigma S_{{\rm min}}=-{\sigma\over 2}\int_0^T\sqrt{(z\cdot\dot{z})^2-z^2\dot{z}^2},
\end{equation}
where $\sigma$ is the string tension, entering, according to the premises of the SVM, through the relation
$\sigma\equiv {1\over 2}\int d^2z D(z^2)$, while $S_{{\rm min}}$ rpresents the minimal surface
bounded by the Wilson loop associated with the closed quark contour. For the particular configuration
given by the straight line segment solution $z^0_\mu$, with reference to our
parametrization of the four-point function loop, one obtains
\begin{equation}
C^{(2)}_{\rm bkgd}[z]\approx -2\sigma\{s_1(s_2-s_1)|q_1\cdot q_2|
+(s_2-s_1)(s_3-s_2)|q_2\cdot q_3|\}
\end{equation}
from which one surmizes that the corresponding corrections to the linear (eikonal) result 
will be $\sim\sigma/s$.

Referring, now. to Eq (14) the first order non-perturbative contribution to the `particle action' reads
\begin{equation}
z_\mu^{{\rm cl}}(t)=z_\mu^0(t)+2\sigma\int_0^Tdt'\Delta(t,t')
\frac{\delta S_{{\rm min}}[z^0]}{\delta z^0_\mu(t')}+{\cal O}(\sigma^2,g^2).
\end{equation}
One determines
\begin{equation}
\frac{\delta S_{{\rm min}}}{\delta z^0_\mu(t')} =\frac{1}{\sqrt{g^{(0)}(t')}}
[z^0(t')\cdot\dot{z}^0(t')\dot{z}_\mu^0(t')-z^0_\mu(t')(\dot{z}^0(t'))^2],
\end{equation}
where
\begin{equation}
g^{(0)} \equiv(z^0\cdot\dot{z}^0)^2-(z^0)^2(\dot{z}^0)^2.
\end{equation}

It is a straightfoward matter to deduce that, for the kinematical region under consideration, the
the intervals $(0,s_1]$ and $(s_3,s_4]$ remain linear while the other two segments are deformed. 
Specifically, one obtains
\begin{equation}
t'\in(0,s_1]:\quad z^{{\rm cl}}_\mu(t)=2i\tilde{q}_{1\mu}t,
\end{equation}
\begin{equation}
t'\in(s_1,s_2]:\quad z^{{\rm cl}}_\mu(t)=2i\tilde{q}_{2\mu}(t-s_1)-2i\bar{q}_{2\mu}\sigma(t-s_1)^2+
z^{{\rm cl}}_\mu(s_1), 
\end{equation}
\begin{equation}
t'\in(s_2,s_3]:\quad z^{{\rm cl}}_\mu(t)=2i\tilde{q}_{3\mu}(t-s_2)-2i\bar{q}_{3\mu}\sigma(t-s_2)^2+
z^{{\rm cl}}_\mu(s_2), 
\end{equation}
and
\begin{equation}
t'\in(s_3,s_4]:\quad z^{{\rm cl}}_\mu(t)=-2i\tilde{q}_{4\mu}(T-t).
\end{equation}
In the above relations the following auxiliary `momentum' variables have been introduced
\begin{equation}
\bar{q}_{2\mu}\equiv q_{2\mu}-\left|\frac{q_2^2}{q_1\cdot q_2}\right|q_{1\mu},
\end{equation}
\begin{equation}
\bar{q}_{3\mu}\equiv q_{3\mu}-\left|\frac{q_3^2}{q_2\cdot q_3}\right|(q_{2\mu}
+\frac{s_1}{s_2-s_1}q_{1\mu})
\end{equation}
while the re-adjusted momenum variables are given by
\begin{equation}
\tilde{q}_{1\mu}\equiv q_{1\mu}+2\sigma\left[\bar{q}_{2\mu}\left(s_2-s_1-\frac{s_2^2-s_1^2}{2T}\right)
+\bar{q}_{3\mu}\left(s_3-s_2-\frac{s_3^2-s_2^2}{2T}\right)\right],
\end{equation}
\begin{equation}
\tilde{q}_{2\mu}\equiv q_{2\mu}+2\sigma\left[\bar{q}_{2\mu}\left(s_2-s_1-\frac{s_2^2-s_1^2}{2T}\right)
+\bar{q}_{3\mu}\left(s_3-s_2-\frac{s_3^2-s_2^2}{2T}\right)\right],
\end{equation}
\begin{equation}
\tilde{q}_{3\mu}\equiv q_{1\mu}+2\sigma\left[-\bar{q}_{2\mu}\frac{s_2^2-s_1^2}{2T}
+\bar{q}_{3\mu}\left(s_3-s_2-\frac{s_3^2-s_2^2}{2T}\right)\right],
\end{equation}
and
\begin{equation}
\tilde{q}_{4\mu}\equiv q_{4\mu}+2\sigma\left[-\bar{q}_{2\mu}\frac{s_2^2-s_1^2}{2T}
-\bar{q}_{3\mu}\frac{s_3^2-s_2^2}{2T}\right].
\end{equation}
The resulting worldline configuration for the four-point (Sudakov) process is depicted in fig 2.

We now proceed to substitute the above `deformed' line paths into the second order expression
for $C^{(2)}_{{\rm pert}}[z]$ given by Eq (20) and asses the new 
worldline configuration associated with the four-point process (for Sudakov kinematics). The
expectation is that the non-linear deformations will produce corrections of the order of
$\sigma/ s$, while the linear ones will simply adjust the corresponding expressions to the 
displaced momentum variables. For the kinematical region under consideration and for $\sigma< m^2$
the results of section 3 remain intact, while for $\sigma>m^2$ we determine
\begin{equation}
\tilde{q}^2_1\sim\sigma{m^2\over s},\quad \tilde{q}^2_2\sim\sigma,\quad
\tilde{q}^2_3\sim s,\quad \tilde{q}^2_4\sim m^2 
\end{equation}
and
\begin{equation}
|\tilde{q}_1\cdot\tilde{q}_2|\sim\sigma,\quad |\tilde{q}_2\cdot\tilde{q}_3|\sim s,\quad
|\tilde{q}_3\cdot\tilde{q}_4|\sim s,\quad |\tilde{q}_4\cdot\tilde{q}_1|\sim\sigma.
\end{equation}
From the above, the following estimates for the quantities $\cosh\gamma_{k,k+1}$ can be deduced
\begin{eqnarray}
&&\frac{|\tilde{q}_1\cdot\tilde{q}_2|}{|\tilde{q}_1||\tilde{q}_2|}\sim\left(\frac{s}{m^2}\right)^{{1\over 2}},\quad
\frac{|\tilde{q}_2\cdot\tilde{q}_3|}{|\tilde{q}_2||\tilde{q}_3|}\sim\left(\frac{s}{\sigma}\right)^{{1\over 2}}
\nonumber\\&&
\frac{|\tilde{q}_3\cdot\tilde{q}_4|}{|\tilde{q}_3||\tilde{q}_4|}\sim\left(\frac{s}{m^2}\right)^{{1\over 2}},\quad
\frac{|\tilde{q}_4\cdot\tilde{q}_1|}{|\tilde{q}_4||\tilde{q}_1|}\sim\left(\frac{s}{m^2}\right)^{{1\over 2}}
\left(\frac{\sigma}{m^2}\right)^{{1\over 2}}.
\end{eqnarray}

Having adjusted our expressions to the new situation, which has brought into 
our analysis input from the non-perturbative (confining) structure of the QCD vacuum, we are ready
to sum the soft effects associated with the four-point Sudakov process. To this end we shall
turn to the renormalization group equation, which will facilitate the running from a low,
`infrared' scale $\mu_{{\rm min}}$ up to the largest energy scale implicated for the 
process, namely $s$. This task will be carried out in the section that follows.

\vspace*{.2cm}

{\bf 5. Renormalization group running}

\vspace*{.1cm}

Repeating the analysis of section 3, one can use the results of the previous subsection to obtain
the modified expressions for the $I_k^{{\rm se,v}},\,k=1,...,4$. We present the
final result for the four-point process, replacing the one given by Eq (44), which reads as follows
\begin{eqnarray}
I^{{\rm se,v}}&\simeq&\frac{\alpha_S}{2\pi}C_F\ln\left(\frac{s}{m^2}\right)\left\{2\ln\left[\frac{\mu^2}
{m\sqrt{\sigma}}\left(\frac{s}{m^2}\right)^{{1\over 4}}\right]+4\ln (Tm\sqrt{\sigma})+\ln(x_1x_2x_3x_4)\right\} 
\nonumber\\&&\quad\quad +\frac{\alpha_S}{2\pi}C_F{1\over 4}\ln^2\left(\frac{\sigma}{m^2}\right).
\end{eqnarray}
The last two terms in the curly brackets are connected with $T$- and $x_k$-integrations, just as 
in the perturbative case. We now introduce the mass scale $\mu_{min}$ 
\begin{equation}
\frac{\partial}{\partial \ln s}I^{{\rm se,v}}(\mu^2=\mu^2_{{\rm min}})=0
\Rightarrow\mu_{{\rm min}}^2=m^2\left(\frac{\sigma}{s}\right)^{1/2}
\end{equation}
to designate the point at which initial conditions for the renormalization group running are
to be set. As noted in the beginning of the present section it should,
of course, hold that $\mu^2_{{\rm min}}>\Lambda_{QCD}^2$. The limit $\sigma\rightarrow m^2$
gives the perturbative estimate  $\mu^2_{{\rm min}}\equiv \frac{m^3}{s^{1/2}}$.

With these specifications let us proceed with renormalization group considerations. Referring to Eq. (6), which gives
the unrenormalized expression for the four-point function, we write, in condensed notation,
\begin{eqnarray}
&&U^{(0)}_{\mu_4\cdot\cdot\cdot\mu_1}\left[\right.\{p_i\},\{s_i\},g]\equiv
\int\limits_{z(0)=z(T)=0}{\cal D}z(t)S_{\mu_4\cdot\cdot\cdot\mu_1}[\dot{z}]
{\rm exp}\left[-{1\over 4}\int_0^Tdt\dot{z}^2(t)-i\sum\limits_{i=1}^{3}p_i\cdot z(s_i)\right]
\nonumber\\&&\quad \times 
\left\langle Tr_C{\rm P} {\rm exp}\left[ig\int_0^Tdt\dot{x}\cdot A\right]\right\rangle_A
= Z[\{p_i\},\{s_i\},\mu,\epsilon]U^{(R)}_{\mu_n\cdot\cdot\cdot\mu_1}[\{p_i\},\{s_i\},g(\mu),\,
\mu],
\end{eqnarray}
where labels $(0)$ and $(R)$ denote unrenormalized and renormalized quantities, respectively. 

We find it convenient to run the quantity
\begin{equation}
\frac{d}{d\ln s}\ln U^{(0)}=\frac{d}{d\ln s}\ln Z +\frac{d}{d \ln s}\ln U^{(R)}
\end{equation}
from $\mu_{{\rm min}}^2$ to $s$. The renormalization 
group equation, which reflects $\mu$ independence, reads 
\begin{equation}
\frac{d}{d{\rm ln}\mu}\frac{d}{{\rm ln}s}{\rm ln}U^{(R)}=
-\frac{d}{d{\rm ln}\mu}\frac{d}{{\rm ln}s}{\rm ln}Z\equiv -2\Gamma
\end{equation}
The anomalous dimension can be read off Eq (82):
\begin{equation}
\Gamma={\alpha_S\over 2\pi}C_F+{\cal O}(\alpha^2_S).
\end{equation}
Retaining leading contributions we present the solution in the form
\begin{equation}
{\rm ln}U^{(R)}(\mu^2=s)=-\int_{m^2}^s{dt\over t}{\rm ln}\left({s\over t}\right)\Gamma[\alpha_S(t)]
-2\int_{\mu_{\rm min}^2}^{m^2}{dt\over t}\ln\left({t\over \mu^2_{{\rm min}}}\right)
\Gamma[\alpha_S(t)]+\ln U^{(R)}(\mu^2=\mu_{{\rm min}}^2),
\end{equation}
which leads to
\begin{equation}
{\rm ln}U^{(R)}(\mu^2=s)=-{4C_F\over\beta_0}\left[ \int\limits_ {{\rm ln}\frac{m^2}{\Lambda^2}}
^{{\rm ln}\frac{s}{\Lambda^2}}\frac{d\xi}{\xi}\left({\rm ln}\frac{s}{\Lambda^2}-\xi\right)
+\int\limits_ {\ln\frac{\mu^2_{{\rm min}}}{\Lambda^2}}
^{{\rm ln}\frac{m^2}{\Lambda^2}}\frac{d\xi}{\xi}\left(\xi-{\rm ln}\frac{\mu_{{\rm min}}}
{\Lambda^2}\right)\right]+\ln U^{(R)}(\mu^2=\mu_{{\rm min}}^2).
\end{equation}

From Eq (74) one determines
\begin{equation}
{\rm ln}U^{(R)}(\mu^2=\mu^2_{{\rm min}})=-{1\over 2\pi}\alpha_S(\mu^2_{{\rm min}})\ln\left({s\over m}\right)
[4\ln(Tm\sqrt{\sigma})+\ln(x_1x_2x_3x_4)]+{\rm sbldg},
\end{equation}
where `sbldg' stands for subleading terms.

If one were to go back to our full expressions and perform the integrations over the
$x_k$ and $T$, one would realize that the coupling $\alpha_S=\alpha_S(\mu^2_{{\rm min}})$
is consistent with the claim $\frac{\alpha_S}{2\pi}C_F\sim \frac{1}{\ln\left(\frac{s}{m^2}\right)}$,
as stated by Eq (46). Accordingly, the nonperturbative contribution re-adjusts the Sudakov 
result for the four-point process as follows
\begin{eqnarray}
{\rm ln}U^{(R)}&=&-\frac{4C_F}{\beta_0} \left[{\rm ln}\left(\frac{s}{\Lambda^2}\right)
{\rm ln}\left(\frac{\ln(s/\Lambda^2)}{\ln(m^2/\Lambda^2)}\right)
-{\rm ln}\frac{s}{m^2}\right.
\nonumber\\&&
+\left.2\ln\frac{m^2}{\mu^2_{{\rm min}}}-2\ln
\left(\frac{\mu^2_{{\rm min}}}{\Lambda^2}\right)
{\rm ln}\left(\frac{\ln(m^2/\Lambda^2)}{{\rm ln}(\mu_{{\rm min}}^2/\Lambda^2)}\right)
\right]+\,{\rm sbldg}.
\end{eqnarray}

Suppose one sets $\alpha_S(s)=\frac{4\pi}{\beta_0}\frac{1}{{\rm ln}(s/\Lambda^2)}$ and keeps terms 
to first order in $\alpha_S$ in the above relation. It would then follow that
\begin{equation}
U^{(R)}\sim \exp\left\{-{\alpha_S\over 2\pi}C_F\left[\ln^2\left({s\over m^2}\right)
+{1\over2}\ln^2\left({s\over \sigma}\right)\right]\right\},
\end{equation}
a result which explicitly displays a visible modification brought about through the inclusion of
non-perturbative input to the analysis of the four-point process. On the other hand,
it is observed that this modification is 
not strong enough to negate the role that Sudakov behavior plays in the analysis of exclusive
processes in QCD.

\vspace*{.3cm}

{\bf 6. Concluding remarks }

\vspace*{.3cm}

In this paper we have defined the Sudakov, QCD mediated, four-point process as one for which: a) A large 
momentum transfer takes place
at each single vertex and b) no observable gluon radiation, in the form of jets, occurs
above a given infrared scale. Such an `idealized' setting allows for a direct study of the
Sudakov suppression in QCD in a way which
is liberated from extra phenomenological burden. The key to atttaining our goal was to
isolate that corner of the kinematical space which realizes the Sudakov conditions
for the process {\it as a whole}. Granted, 
in a realistic situation, such as one pertaining to meson-meson elastic scattering, the 
corresponding kinematical analysis is far more complicated. On the other hand, our 
`idealized' setting is in position to capture the essence of the Sudakov kinematical constraints and
lead to results which check directly with the ones corresponding to the more `realistic' analysis.
In short, our claim is that the corner of the kinematical space identified in this
work corresponds to the one which defines the Feynman picture, as opposed to the quark counting rule, 
for hadron-hadron elastic scattering. The clear dividents of the present approach
is that it enabled us to get a glimpse on non-perturbative alterations to the Sudakov result, as
one approaches $\Lambda_{QCD}$ from above, through the employment of the SVM. 

The following comment, given in retrospect, might be useful. It concerns the utilization of the scale $m^2$ 
for isolating the relevant integration region for Sudakov kinematics, cf. Eq (42). The underlying
hypothesis is that in a complete computation, i.e. with the nonperturbative
contributions fully taken into account, such a scale would be determinable in terms of the
dimensional parameters entering the problem: $m^2=m^2(M^2,\sigma,\cdot\cdot\cdot)$. The nature
of the present calculation was such that it renders the non-perturbative corrections in the region
$m^2>\sigma$ negligible, whereas they become more important for $m^2<\sigma$, cf Eq (84). The point is that
our present effort has contained itself to the perturbative domain of the theory, {\it albeit}
bringing in some non-perturbative input (interference) through the SVM. We anticipate that the extension 
of the present analysis to the Regge kinamatical region (always for the four-point function)
will present richer features due to the fact that the relevant gluon exchanges will occur among
opposite, so to say, (world)line segments, i.e. segments separated by two consecutive vertices.  
Our next goal
is, indeed, to extend the analysis of this work, for the same process, to the Regge kinematical regime.

\vspace*{0.5cm}

\begin{center}
{\bf Acknowledgements}
\end{center}\thispagestyle{empty}

\vspace{0.1cm}

The authors are grateful to Professor Y. A. Simonov for the many critical 
remarks he has made during the course of this work. They also wish to acknowledge the support by the General 
Secretariat of Research and Technology of the University of Athens.

\newpage

\appendix
\setcounter{section}{0}
\addtocounter{section}{1}
\section*{Appendix}
\setcounter{equation}{0}
\renewcommand{\theequation}{\thesection.\arabic{equation}}

In this Appendix we derive Eqs (26)-(28) in the text, starting from Eq. (24). The latter 
becomes, via the use of Eq (18),
\begin{equation}
I_{k-1\,\,\,\,k+1}^{\quad k}={1\over2}g^2C_F\frac{(4\mu^2)^{2-D/2}}{4\pi^{D/2}}\Gamma
\left({D\over 2}-1\right)q_{k+1}\cdot q_k\int_{s_{k-1}}^{s_k}dt_1\int_{s_{k}}^{s_{k+1}}dt_2
\frac{1}{|\sum\limits_{i=1}^3 p_i\left(\Delta(t_2,s_i)-\Delta(t_1,s_i)\right)|^{D-2}}.
\end{equation}
The denominator of the expression inside the integral can be simplified,
once appealing to Eq (18):
\begin{equation}
\sum\limits_{i=1}^{3} p_{i\mu}\left(\Delta(t_2,s_i)-\Delta(t_1,s_i)\right)
=q_{k+1,\mu}(t_2-t_1)+p_{k\mu}(s_k-t_1)
\end{equation}
for $t_2\in[s_{k+1},s_k)$, $t_1\in[s_k,s_{k-i})$

One thereby obtains
\begin{equation}
I_{k-1\,\,\,\,k+1}^{\quad k}={1\over2}g^2C_F\frac{(4\mu^2)^{2-D/2}}{4\pi^{D/2}}\Gamma
\left({D\over 2}-1\right)q_{k+1}\cdot q_k\int_{s_{k-1}}^{s_k}dt_1\int_{s_{k}}^{s_{k+1}}dt_2
\frac{1}{|q_{k+1}(t_2-t_1)+p_k(s_k-t_1)|^{D-2}}.
\end{equation}

Reparametrizing according to $s_k-t_1\rightarrow t_1'$, $t_2-s_k\rightarrow t_2'$ and 
noticing that $q_{k+1}+p_k=q_k$ one finds
\begin{eqnarray}
I_{k-1\,\,\,\,k+1}^{\quad k}&=&{1\over2}g^2C_F\frac{(4\mu)^{2-D/2}}{4\pi^{D/2}}\Gamma
\left({D\over 2}-1\right)q_{k+1}\cdot q_k\int_0^{s_k-s_{k-1}}dt_2\int_0^{s_{k+1}-s_k}dt_1
\frac{1}{|q_{k+1}t_2+q_k t_1|^{D-2}}\nonumber\\&&\quad
={1\over2}g^2c_F\frac{\mu^{4-D}}{4\pi^{D/2}}\Gamma\left({D\over 2}-1\right){2\over D-3}
(\mu^2L_k^2)^{2-D/2}V_D(c_{k,k+1}){1\over 4-D}
\end{eqnarray}
with the $c_{k,k+1}$ as defined in Eq (25), to be written simply as $c$ from hereon.

In the above result we have introduced
\begin{equation}
V_D\equiv{1\over 2}A_D(x,c)+{1\over 2}x^{4-D}A_D({1\over x},c),
\end{equation}
where
\begin{equation}
x\equiv \frac{L_{k+1}}{L_k}=\frac{|q_{k+1}|\,|s_{k+1}-s_k|}{|q_{k}|\,|s_k-s_{k-1}|}
\end{equation}
and
\begin{equation}
A_D(x,c)\equiv c^2F\left(1,{D-2\over2};{D-1\over2};1-c^2\right)-c\frac{x+c}{(x^2+2xc+1)^{D/2-1}}
F\left(1,{D-2\over2};{D-1\over2};\frac{1-c^2}{x^2+2xc+1}\right).
\end{equation}
For $D=4$ one obtains
\begin{equation}
A_4(x,c)= \frac{c}{\sqrt{1-c^2}}\arctan \frac{\sqrt{1-c^2}}{c}-\frac{c}{\sqrt{1-c^2}}
\arctan\frac{\sqrt{1-c^2}}{x+c},
\end{equation}
which, upon substitution in Eq (A5), leads to the desired result.

\newpage

\begin{center}

{\bf Figure captions}

\end{center}

\vspace{.2cm}

Fig. 1. The four-point process under cosideration in the eikonal approximation (straight worldline segments).

\vspace{.2cm}

Fig. 2. Deformed four-point worldline contour on the account of SVM induced non-perturbative effects
(still, slightly above $\Lambda_{QCD}$).


\begin{thebibliography}{99}


\bibitem{1} V. Sudakov, Sov. Phys. JETP {\bf 3}, 65(1956)

\bibitem{2} J. Botts and G. Sterman, Nucl. Phys. {\bf B325}, 63(1989).

\bibitem{3} A. H. Mueller, Phys. Rep. {\bf 73}, 237(1981).

\bibitem{4}  G. P. Korchemsky and G. Sterman, Nucl. Phys. {\bf B437}, 415(1995).

\bibitem{5} V. A. Fock, Izvestiya Akad. Nauk. USSR, OMEN, p. 557(1937).

\bibitem{6} R. P. Feynman, Phys. Rev. {\bf 80}, 440(1950).

\bibitem{7} J. Schwinger,  Phys. Rev. {\bf 82}, 664(1951).

\bibitem{Strassler:1992zr} M.~J.~Strassler, 
Nucl.\ Phys.\ {\bf B385} (1992)145.  

\bibitem{Reuter:1997zm} M.~Reuter, M.~G.~Schmidt and C.~Schubert, 
Annals Phys.\ {\bf 259} (1997)313. 

\bibitem{10} A.~I.~Karanikas and C.~N.~Ktorides, Phys Lett {\bf B331}, 403(1992); Phys. Rev 
{\bf D52}, 5883(1995).

\bibitem{11} A. I. Karanikas, C. N. Ktorides and N. G. Stefanis, Eur. Phys. J. {\bf C26}, 445(2003);
G. C. Gellas, A. I. Karanikas and C. N. Ktorides, Annals Phys. (NY) {\bf 255}, 228(1997). 

\bibitem{12}H. G. Dosch, Phys. Lett. {\bf B190}, 177(1987).

\bibitem{13} H. G. Dosch and Yu. A Simonov, Phys. Lett. {\bf B205}, 339(1988).

\bibitem{14} Yu. A. Simonov, Nucl. Phys. {\bf B307}, 512(1988).

\bibitem{16} Yu. A. Simonov, Yad. Fiz. {\bf 54}, 192(1991).

\bibitem{17} A. Di Giacomo, H.G. Dosch, V.I. Shevchenko and Yu.A. Simonov, Phys.Rept. {\bf 372}(2002).

\bibitem{18} A. I. Karanikas and C. N. Ktorides, JHEP {\bf 9911}, 033(1999).

\bibitem{Polyakov:sk} A.~M.~Polyakov in: E.~.~Brezin and J.~.~Zinn-Justin,  
``Fields, Strings And Critical Phenomena'' {\it Amsterdam, Netherlands: North-Holland (1990)}. 


\bibitem{20} N. G. Van Kampen, Physica {\bf 74} 215, 239(1974); Phys. Rep. C {\bf 24}, 172 (1976).

\bibitem{21}  O. Nachtmann (Heidelberg U.),
Lectures given at the 35th International University School of Nuclear 
and Particle Physics: Perturbative and Nonperturbative Aspects of Quantum Field Theory, 
Schladming, Austria (1996) HD-THEO-96-38.

\bibitem{27} G. K. Savvidi, Phys. Lett. B {\bf 71}, 133(1977).

\bibitem{18} Yu. A. Simonov, in {\it Lecture Notes in Physics} {\bf 479}, Springer 144(1995).

\bibitem {'tHooft:1975vy} G.~ 't Hooft, 
``The background field method in gauge field theories'',
in *Karpacz 1975, Proceedings, Acta Universitatis Wratislaviensis No.368, Vol.1, 345 
(Wroclaw 1976).

\bibitem{29} V. A. Fock, Sov. Phys. {\bf 12}, 684(1937).


\end{thebibliography}
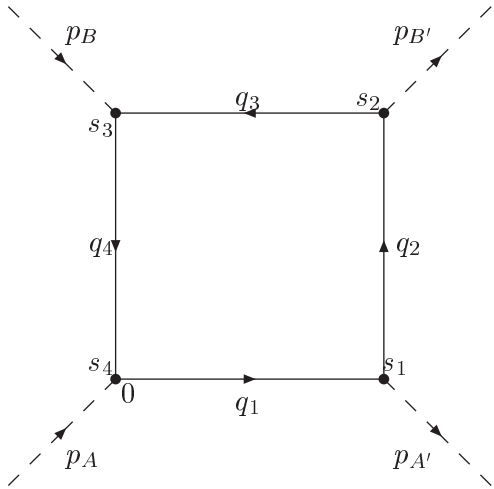
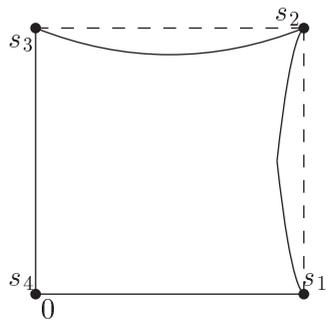
\end{document}